\documentclass[a4paper,12pt,superscriptaddress,showkeys]{revtex4}

\usepackage[centertags]{amsmath}
\usepackage{amsfonts}
\usepackage{amssymb}
\usepackage{newlfont}
\usepackage{stmaryrd}
\usepackage{mathrsfs}
\usepackage{euscript}
\usepackage{graphicx}

\usepackage{color}

 \let\be=\beta  \let\ep=\epsilon
   
 \let\la=\lambda \let\om=\omega 
  \let\th=\theta

\let\De=\Delta  \let\La=\Lambda

\newcommand{\caT}{{\mathcal T}}

\newcommand{\bbP}{{\mathbb P}}

\newcommand{\opunit}{\text{1}\kern-0.22em\text{l}}

\DeclareMathAlphabet{\mathpzc}{OT1}{pzc}{m}{it}

\newcommand{\rel}{\,|\,}

\newcommand{\id}{\textrm{d}}

\begin{document}

\title{\large{{\bf Dynamical fluctuations for semi-Markov processes}}}

\author{Christian Maes}
\email{christian.maes@fys.kuleuven.be}
\affiliation{Instituut voor Theoretische Fysica, K.U.Leuven, Belgium}
\author{Karel Neto\v{c}n\'{y}}
\email{netocny@fzu.cz}
\affiliation{Institute of Physics AS CR, Prague, Czech Republic}
\author{Bram Wynants}
\email{bram.wynants@fys.kuleuven.be}
\affiliation{Instituut voor Theoretische Fysica, K.U.Leuven, Belgium}

\begin{abstract}
We develop an Onsager-Machlup-type theory for nonequilibrium semi-Markov processes.
 Our main result is an exact large time
asymptotics for the joint probability of the occupation times and the currents in the system, establishing some generic large deviation structures. We discuss in detail how the
nonequilibrium driving and the non-exponential waiting time distribution influence
the occupation-current statistics. The violation of the Markov condition is reflected in the emergence of a new type of nonlocality in the fluctuations.  Explicit solutions are obtained for some examples of driven random walks on the ring.
\end{abstract}


\maketitle
\section{Introduction}

Stochastic processes enter physics because of some reduced or
incomplete description in terms of variables whose states at
earlier times do not uniquely determine their future states.
Unless we go to infinite scale separations and treat the relevant
set of variables as for example in the hydrodynamic limit, we
expect that the reduced description allows for fluctuations and
randomness in the dynamics. We then speak about the mesoscopic
level of description for which reproducibility which is so typical
for the macroscopic world, is not available yet.

Furthermore, the resulting or effective descriptions in terms
of stochastic dynamics cannot always be reliably treated within the
Markov approximation.  That Markovian level would also require
particular time-scale separations such that the random evolution
becomes essentially memoryless. One easily looses the Markov
property when the landscape of states and their mutual connections
are getting very complicated and some further coarse-graining
collects various states into one, if only to simplify things. For
example in the theory of spin glasses, by combining various local
minima of the free energy in one state and depending on the various
activation energies, one could effectively obtain stretched
exponential waiting time distributions.  Or, for transport in
strongly disordered systems, one can think that conduction is the
result of a large series of hops, and the effective or total
hopping rate must convolute different exponentials, cf.~Mott's
variable range hopping~\cite{mot}. For biophysical processes such
as in molecular motors or in ion channels one observes gate states
that are either at the beginning or at the end of a series of
states that appear different only in some minor (internal)
rearrangement of molecules.  The passage from entrance to exit
through these internal and largely hidden states can again give
rise to nonexponential waiting time distributions. See
\cite{Ku,wq,qw} for specific biomolecular realizations. In the
examples that follow in section \ref{exa}, we present a very
simple scenario of such a transport problem. The natural
probabilistic environment is then that of semi-Markov processes,
\cite{mon,kf}. Other scenario's can be due to the random nature of
energy levels such as conjectured in blinking quantum dots for
which the phenomenology suggests
power law waiting time distributions, \cite{BQD}.\\

The purpose of the
present paper is to investigate the role of entropy fluxes and of
dynamical activity in the fluctuation theory for semi-Markov processes.
  Such a theory has been investigated for
Markov jump and Markov  diffusion processes in the light of recent
studies in nonequilibrium statistical mechanics,
\cite{ber,der,bern,MN}. It remains important to characterize the
fluctuation functionals in their physical role away from the
strict Markovian context.  Going now beyond these Markov processes,
we are especially interested in the
influence and in the role of the waiting time distribution in the
dynamical fluctuations.  After all, it parameterizes a
time-symmetric factor in the transition events, whose influence on
the joint occupation and current statistics needs to be
understood. In this sense, we go here also beyond previous approaches 
(e.g.~\cite{ag}), where only the time-antisymmetric factor is considered.
Our results give detailed expressions for the
fluctuation functionals and we interpret the role of the waiting
time distribution in them. We provide some more results in the regime of small fluctuations,
where we prove that the current and occupation fluctuations decouple close to equilibrium.
Here we also conclude that occupation fluctuations are in a sense more sensitive to non-Markovian behaviour
than current fluctuations.\\

The next section contains  a brief introduction to the world (and
the notation) of semi-Markov processes. Further elements of the
semi-Markov theory are recalled in the Appendix.  We then move to
a general introduction on dynamical fluctuation theory, which
contains our main formulations of the fluctuation functionals for
semi-Markov processes.  Subsections are devoted to some
corollaries and to the interpretation of these functionals. That
includes a fluctuation symmetry for the entropy production and in section \ref{small} the
treatment of small fluctuations around a nonequilibrium state. We
end in Section \ref{exa} with examples of semi-Markovian transport
in rings for which our results are getting fully explicit (for small fluctuations).
In a more specific example we also compute the generating function and the
first few cumulants for the current fluctuations.

\section{Semi-Markov process}\label{sm}

\subsection{Definitions and notation}

We consider jump processes on a finite space $\Omega$ with states
denoted by $x,y,\ldots$ The updating is time-homogeneous and in
continuous time.  Semi-Markov processes are non-Poissonian with a
renewal property. This means that the probability of a jump from
$x$ to $y$ at a certain time depends only on the states $x$, $y$ and
the time $t$ since the last jump occurred. More precisely,
let us denote $Q(x,y,t)$ for the density of random
transitions at time $t$ from the state
$x$ to $y$. This so called semi-Markov kernel defines the process.
Further it is useful to introduce
\begin{equation}\label{defs}
  Q(x;t) = \sum_yQ(x,y;t), \qquad
  p(x,y) =\int_0^{+\infty} Q(x,y;t)\,\id t
\end{equation}
which are respectively the waiting (or sojourn) time distribution
in $x$ and the transition probabilities regardless of the waiting time, and
\begin{equation}
  \La(x; t) = \int_t^{+\infty} Q(x; \tau)\,\id\tau
\end{equation}
the probability that the system rests at state $x$ for at least time $t$.
In the sequel we always assume that
$Q(x,y;t) = O(t^{-3-\ep})$, $\ep > 0$ asymptotically for $t \to +\infty$ so
that the first and the second moments with respect to the distributions
$Q(x,y;t)$ are finite. An important role plays the effective escape rate
$\bar\la(x)$ from $x$ defined as the reciprocal to the average sojourn time
in $x$, i.e.,
\begin{equation}\label{lbar}
 \frac{1}{\bar{\lambda}(x)} =
 \int_0^{+\infty} \tau\,Q(x;\tau)\,\id \tau
  = \int_0^{+\infty} \La(x;\tau)\,\id\tau
\end{equation}
We say that the semi-Markov process enjoys time-direction independence when
\begin{equation}\label{tdi}
Q(x,y;t) = p(x,y)\,Q(x;t)
\end{equation}
That is the case when the waiting time only depends on the present
(and not on the future) state.  We make that assumption
\eqref{tdi} throughout the paper. See section \eqref{ex1} and
\eqref{1ex} for a
specific (counter)example.

If the process is Markov then
\[
  Q(x,y;t) = p(x,y)\,\lambda(x)\,e^{-\lambda(x)t}\,,\qquad
  \La(x; t) = e^{-\la(x) t}
\]
with $\lambda(x)$ the escape rate from state $x$;
the product $w(x,y) =\lambda(x)\,p(x,y)$ is called the transition rate.
The effective escape rates are $\bar{\lambda}(x) = \lambda(x)$.\\

Further elements of the theory of
semi-Markov processes are summarized in the Appendix.  In
particular, there we review a derivation of the (generalized) Master equation, and its
formulation from the point of view of an embedded Markov chain.
More details are of course available in the literature, see
e.g.~\cite{mon,qw} for physics introductions.

\subsection{Semi-Markov statistics}

We add here some ingredients of the theory of semi-Markov processes that relate
to the statistical mechanics we are going for in the next section.\\

As explained in the Appendix, see~\eqref{genm} and Section~\ref{sta},
the stationary distribution $\rho$ solves the stationarity equation
\begin{equation}\label{eq: stat-cond}
\begin{split}
  \sum_y j_{\rho}(x,y) = 0\,,\qquad
  j_\rho(x,y) =
  \rho(x)\bar{\lambda}(x)p(x,y) - \rho(y)\bar{\lambda}(y)p(y,x)
\end{split}
\end{equation}
where $j_\rho(x,y)$ is the stationary (probability) current from state
$x$ to $y$. Remark that this equation coincides with the stationarity
condition for a continuous time Markov process with transition rates
$\bar{\lambda}(x) p(x,y)$. In particular, the stationary distribution
and currents depend on the waiting time distributions only through the
effective escape rates.\\

In this paper we go beyond the above stationary characterization of the
semi-Markov process and we want to understand the structure of fluctuations
of the occupations and currents around their stationary values, i.e., to develop
a dynamical fluctuation theory for these processes. For that we need more
details about the process and a natural starting point is the path-space
distribution evaluating the plausibility of trajectories. This was the general
idea of Onsager and Machlup, \cite{ons}. Here we follow the strategy developed
in~\cite{mnw} for Markov systems.\\

A path $\omega = (x_t)_{0\leq t\leq T}$ specifies the sequence of states
together with the jump times,
\begin{equation}\label{eq: path}
  (x_0,t_1,x_1,t_2,\ldots,x_{n-1},t_n,x_n)\,, \quad
  0\leq t_1 < t_2<\ldots t_n\leq T\,, \quad
  n = 1,2,\ldots
\end{equation}
In order to construct a transient semi-Markov process started from a
given initial distribution $\mu$ at time zero, we make a ``stationarity''
assumption about the history of the process in negative times: we let the
age of an initial configuration sampled from $\mu$ be random and conditionally
distributed according to the stationary process. Then the resulting path space
distribution $P_{\mu}(\omega)$ giving the probability of a path $\omega$
has the density
\begin{multline}\label{path}
  \id P_{\mu}(\omega) =
  \mu(x_0)\,\bar{\lambda}(x_0)\,\Lambda(x_0,x_1;t_1)\,
  Q(x_1,x_2;t_2-t_1)\ldots
\\
  \ldots Q(x_{n-1},x_n;t_n-t_{n-1})\,\Lambda(x_n;T-t_n)\,
  \id t_1\ldots \id t_n
\end{multline}
in which $\Lambda(x_0,x_1;t_1) = \int_{t_1}^{+\infty}Q(x_0,x_1;\tau)\,\id\tau$
is the waiting time distribution for the initial interval
$[0,t_1]$, taking into account the random (stationarily distributed) age of
the initial configuration $x_0$ at time zero when the process starts.
Similarly, the last term, $\La(x_n; T - t_n)$, comes out by integrating
the waiting time distribution $Q(x_n; t_{n+1})$ over all possible times of
the first jump outside the time interval, $t_{n+1} > T$. Note that for
$\mu = \rho$ the above construction yields a stationary process.

The apparent similarity between the first and the last terms in~\eqref{path},
representing the past and the future of the process, will be exploited next
in the analysis of the time-reversal symmetry and its breaking.

\subsection{Time-reversal and local detailed balance}

Any open system weakly coupled to its environment and being in thermal equilibrium 
with the latter has to satisfy two general conditions that directly follow from first 
principles: (i) its stationary distribution has the canonical form
\begin{equation}\label{eq: canonical}
  \rho(x) = \frac{1}{Z}\,e^{-\beta U(x)} 
\end{equation}  
with $U$ the energy of the system and $\be$ the bath temperature, and (ii)
the (effective) stochastic dynamics of the system is symmetric under time-reversal. 
This symmetry can be broken either by  starting the dynamics from a nonstationary 
condition (i.e., in transients), or provided the system is coupled to several thermal 
reservoirs mutually not in thermal equilibrium (i.e., in transport processes).
Then the dynamics of the system is no longer time-reversal symmetric, however, this 
symmetry is broken in a very specific way: a natural `measure' of irreversibility, 
see below, coincides with the change of entropy together in the system and in the 
environment. For a general argument see, e.g., \cite{mnt}. Next we specify these 
considerations to the semi-Markov processes.\\

As a standard measure of irreversibility of the process we consider
the path-dependent quantity defined as the relative plausibility of a path
with respect to its time-reversed counterpart. 
Introducing the time-reversal $\th\om$ of a path $\om$ as
$(\th\om)_t = \om_{T - t}$ or
\begin{equation}
  \th\om = (x_n, T - t_n, x_{n - 1}, T - t_{n-1},\ldots, x_1, T - t_1, x_0)
\end{equation}
cf.~\eqref{eq: path}, we define
\begin{equation}\label{pathentropy}
\begin{split}
  S_{\mu}(\omega) &=
  \log\frac{\id P_{\mu}(\omega)}{\id P_{\mu_T}(\theta\omega)}
\\
  &= \log\frac{\mu(x_0)\bar{\lambda}(x_0)\Lambda(x_0,x_1;t_1)\ldots
  Q(x_{n-1},x_n;t_n-t_{n-1})\Lambda(x_n;T-t_n)}{\mu_T(x_n)\bar{\lambda}(x_n)
  \Lambda(x_n,x_{n-1};T-t_n)\ldots Q(x_{1},x_0;t_2-t_{1})\Lambda(x_0;t_1)}
\end{split}
\end{equation}
where $\mu_T$ is the distribution at time $T$ as evolved from the
$\mu$ at time zero, i.e., the solution of the generalized Master equation~\eqref{genm} or~\eqref{malap}.
As we restrict ourselves to the case of time-direction independence
\eqref{tdi}, equation (\ref{pathentropy}) considerably simplifies
to
\begin{equation}\label{8}
  S_{\mu}(\omega)  =   \log \mu(x_0) - \log \mu_T(x_n) +
  \sum_{i=1}^n \log \frac{\bar{\lambda}(x_{i-1})p(x_{i-1},x_i)}
  {\bar{\lambda}(x_i)p(x_i,x_{i-1})}
\end{equation}
which is formally the same as one has for a Markov process with escape rates $\bar{\lambda}(x)$. 

Assume first that the system is in thermal equilibrium with a heat bath. 
Then $\mu = \rho$ is given by~\eqref{eq: canonical} and, by time-reversibility,
$S_\rho(\om) = 0$, pathwise. This is equivalent to
\begin{equation}\label{detbal}
  \log \frac{\bar{\lambda}(x)p(x,y)}{\bar{\lambda}(y)p(y,x)} =
  \beta\, [U(x) - U(y)]
\end{equation}
which is a generalized detailed balance condition (recall that in the Markov case, 
$w(x,y) = \la(x)\,p(x,y)$ are transition rates). Using~\eqref{eq: stat-cond}, 
this is further equivalent to the absence of all stationary currents, 
$j_\rho(x,y) = 0$. A rigorous argument for the equivalence between the (generalized) 
detailed balance condition and the time-reversibility of time-direction independent 
semi-Markov processes can be found in~\cite{cha,wq}. Remark that the assumption of 
time-direction independence is crucial here and cannot be easily abandoned. 

To make a step beyond thermal equilibrium, observe first that the right-hand side 
of~\eqref{detbal} reads $\beta$ times the heat flux (= entropy flux) into the heat 
bath, per a single transition $x \longrightarrow y$ in the system. This is clearly 
a global condition since all the local entropy fluxes derive from a potential 
(or state quantity) $\be U$. However, it has a natural local variant that only requires that
\begin{equation}\label{ldb}
  \log \frac{\bar{\lambda}(x)p(x,y)}{\bar{\lambda}(y)p(y,x)} =
  \text{entropy flux } (x\rightarrow y)
\end{equation}
no matter whether the entropy fluxes derive from a potential or not. Physically this 
corresponds to a system coupled to several heat baths on nonequal temperatures, assuming 
that each transition $x \longrightarrow y$ is assisted by no more than one reservoir. 
The condition~\eqref{ldb} is called local detailed balance. When modeling a particular 
physical process,
we usually take the individual entropy fluxes per each transition as \emph{a priori} known, cf.~\cite{bl}.

Under the local detailed balance condition, the path quantity $S_\mu$ is the sum of two terms: of
the difference $-\log\mu_T(x_n) + \log\mu(x_0)$ which is to be understood as the variable entropy increase
in the system (one checks that its expectation equals the increase in Shannon entropy), and of the total entropy
flux into environment. The latter adds up contributions from all transitions along the random trajectory.\\

Let us conclude here with two remarks: first of all, local detailed balance
is not a mathematical condition but rather a general guiding principle
to be followed when modeling an arbitrary open system driven out of equilibrium.
It is mainly because of this that we restrict ourselves to time-direction independent processes.
For time-direction dependent processes the assumption of local detailed balance does not
seem to make sense.
Furthermore, a direct consequence of local detailed balance is a
symmetry in the fluctuations of the time-integrated entropy flux
and also of the time-integrated currents (fluctuation theorems). It follows easily from
\eqref{8} via standard manipulations. See also~\cite{ag,ell,mnt}
for another and more detailed approach.

\subsection{Resolution with respect to time-reversal}

By construction, the entropy flux is intimately related to the time-antisymmetric part 
in the logarithmic probabilities (or action). In order to separate more
explicitly the time-symmetric sector of fluctuations from the time-antisymmetric
one, we will make a parametrization, inspired by the Markov case~\cite{mnw}: there 
one relates the transition rates $k(x,y)$ to an equilibrium reference process with rates $k_0(x,y)$ so that
\[ k(x,y) = k_0(x,y)\,e^{\frac{1}{2}\,G(x,y)} \]
and $G(x,y) = -G(y,x)$. (One checks that for a Markov processes such a representation always exists). 
By local detailed balance, $G(x,y)$ can be seen as ($\be$ times) the work done by an extra (with 
respect to the reference) force along the transition $x\to y$.

As a generalization, we consider as a suitable equilibrium reference
system another semi-Markov process with the waiting time distributions
$Q_0(x,y;t) = Q_0(x;t) p_0(x,y)$ such that
$\log\, [p_0(x,y) / p_0(y,x)]$ derives from a potential (the global detailed balance
 condition) and that the waiting times of the original and the reference processes are related by~\cite{remark1}
\begin{equation}\label{compare}
 Q(x,y;\tau) = Q_0(x,y;\tau)\,e^{\frac{1}{2}\,G(x,y)+\Delta^0_{G}(x)\tau}
\end{equation}
for some $G(x,y)  = -G(y,x)$. The term $\Delta^0_{G}(x)$ is a compensator fixed by the normalization condition
\begin{equation}\label{deltacond}
  \int_0^{\infty}\id \tau\, Q(x;\tau) = 1
\end{equation}
Note that the parametrization~\eqref{compare} preserves the time-direction independence property.
By comparing with condition~\eqref{ldb} of local detailed balance, the entropy flux per transition
$x \longrightarrow y$ reads
\begin{equation}\label{ldbref}
  \log \frac{\bar{\lambda}(x)p(x,y)}{\bar{\lambda}(y)p(y,x)} =
  G(x,y) + u(x) - u(y)
\end{equation}
with some potential $u$ that is explicitly computable.
The process breaks the (global) detailed balance unless $G(x,y)$ also derives from a potential.
We can therefore say that $G(x,y)$ is the forcing of the process. The decomposition of the entropy
flux into potential and nonpotential parts is hence fixed by comparing to a particularly chosen
reference equilibrium.

Here is how we split the action
into, respectively, a time-antisymmetric and a time-symmetric
parts. From~\eqref{path}, the logarithmic density of our process
with respect to the equilibrium reference process $P^0_\mu$ is
\begin{equation}\label{ks}
  \log \frac{\id P_{\mu}(\omega)}{\id P_{\mu}^0(\omega)}
  \doteq \frac{1}{2}\sum_{t\leq T}G(x_{t-},x_{t+})
  +\int_0^{T}\id t\,\Delta_G^0(x_{t})
\end{equation}
where $\doteq$ denotes that we have only taken the time-extensive part,
and neglected temporal boundary terms that will become redundant.
In the first term the
sum is over all jump times in
$\omega$ and by \eqref{ldbref} it is equal to the total (i.e., time-integrated) entropy flux
along path $\om$, for the original process. The second term in~\eqref{ks} is manifestly
time-symmetric and it can be understood as a time-undirected dynamical activity,
or what we have called traffic in~\cite{MN,mnw}
(relatively with respect to the reference dynamics.)

The path-space average of~\eqref{ks} with respect
to our process gives its dynamical or also called,
Kolmogorov-Sinai entropy---the relative
entropy between the process and its reference as distributions on
paths. The decomposition~\eqref{ks} suggests to define two
functionals which take averages over the two terms in~\eqref{ks} separately.
They will appear later; for the stationary regime with stationary
density $\rho$ we divide \eqref{ks} by $T$ and let
$T\to +\infty$ to write
\[
  \frac{1}{T} \Bigl\langle\, \log \frac{dP_{\rho}}{dP_{\rho}^0}\,
  \Bigr\rangle \longrightarrow \frac{\dot{S}}{2}  + \dot{\caT}
\]
with
\begin{eqnarray}\label{fun}
 \dot{S} &=& \sum_{x,y} \rho(x)\,\bar{\lambda}(x)\,p(x,y)\,G(x,y)
  = \frac {1}{2}\sum_{x,y}j_\rho(x,y)\,G(x,y)\nonumber  \\
 \dot{\caT}&=& \sum_x\rho(x)\,\Delta_G^0(x)
\end{eqnarray}
The quantity $\dot S$ is the stationary average of the entropy flux per unit time.
Note that it does not depend on the particular choice of the reference equilibrium
process as it is insensitive to adding any potential difference to the driving $G$.
The second component of the dynamical entropy, $\dot\caT$, measures the stationary
dynamical activity in the sense of its ``excess'' with respect to the equilibrium
dynamics taken as a reference.

\section{Dynamical fluctuations}\label{fluct}

\subsection{Scope}
Equilibrium statistical mechanics provides us with a fluctuation
theory through which the thermodynamic potentials can also be
understood as fluctuation functionals. Natural variables for these functionals usually
are the energy and particle densities, the magnetization or still other characteristics
of an equilibrium state. The precise formulation of all that is found in the theory of
large deviations as pioneered by Boltzmann, Planck, and Einstein, which starts from the
identification of the thermodynamic entropy with the logarithm of a probability, see
e.g.~\cite{lan,mar,el}. We call this a static fluctuation theory where the main extensive
parameter is the spatial volume or the number of particles.
That also has an extension to spatially extended systems out of equilibrium but there is 
no simple way of determining the stationary distribution. The difficulties
with its direct determination can be overcome by analyzing typical paths along which
 macroscopic fluctuations get spontaneously created, via exploiting methods of
analytical mechanics. The fluctuation functionals, often called nonequilibrium
free energies, are then found to solve an appropriate Hamilton-Jacobi equation \cite{jo}.

In contrast, dynamical fluctuation theory deals with deviations from stationary behavior
that are observed over a large time period. This formulation is
most useful for mesoscopic systems in contact with large external
reservoirs, where these fluctuations can be visible on the level
of the system. An immediate consequence is the variational
characterizations of the steady state regime, much as the Gibbs
variational principle characterizes thermal equilibrium from the
minimum of a free energy functional. Beyond that, the question is
once more whether the corresponding (now, dynamical) fluctuation functionals
allow for a natural physical interpretation, whether they can provide relations
between quantities directly accessible via measurement etc.
Indeed, we recall that in equilibrium the Onsager-Machlup theory constructs actions
for the distribution of macroscopic histories that relate to response coefficients
and to dissipation functions, \cite{ons}. It would be most interesting to
obtain an extension of these functionals to reach domains further away from
equilibrium and also in situations different from those of fluctuating hydrodynamics.

Here we consider the set-up of semi-Markovian jump processes; the Markov case has
been discussed in \cite{mnw}.  The questions can however be put in a more general
context, as now follows.\\

We consider a path (or history or trajectory) and we observe the
occupation of states and  the various transitions over states in
some large time-interval $[0,T]$.  More precisely, we first look
at the fraction of time that the system spends in a state $x\in
\Omega$ for one specific path $\omega$:
\begin{equation}\label{occ}
 \mu_T(x):= \frac{1}{T}\int_0^T \delta[x_t=x]\,\id t
\end{equation}
where  $\delta[x_t=x]$ is $1$ whenever $x_t=x$ and zero otherwise.
Obviously, $\mu_T(x), x\in \Omega$, defines a probability law but
it is itself random as dependent on the stochastic trajectory
$\omega$. We assume that these paths are drawn from the unique
steady state with stationary density $\rho$. Then, as time
$T\to +\infty$ we have convergence of $\mu_T(x)$ to that
$\rho(x)$; that corresponds to an assumption of ergodicity.
Secondly we define the empirical densities of jumps between
states $x$ and $y$:
\begin{equation}
 k_T(x,y)
 := \frac{1}{T}\sum_{t\leq T} \delta[x_{t-}=x]\,\delta[x_{t+} = y]
\end{equation}
where  the sum is over all jump times and
$x_{t\mp}$ are the configurations before and after the jump, respectively.
Again, that is a random quantity, typically converging for large $T$ to
$\rho(x) \bar{\lambda}(x) p(x,y)$. Finally there is the empirical
current
\begin{equation}
 j_T(x,y)=k_T(x,y) - k_T(y,x)
\end{equation}
The question of our dynamical fluctuation theory is to see and to
physically understand the asymptotic statistics for $\mu_T(x)$ and
$j_T(x,y)$. What values do these assume and with what probability?
In other words, we take a probability law $\mu$ on $\Omega$ and a
family $j = (j(x,y))$ with $\sum_y j(x,y) = 0$ and we ask for the
steady state probability
\begin{equation}\label{dv}
\bbP[\mu_T \simeq \mu\,;\, j_T \simeq j] \propto e^{-T\,I(\mu,j)}
\end{equation}
as $T\to +\infty$, to realize these $\mu$ and $j$ along the
trajectories.  We already suggest here that there exists a rate
function $I(\mu,j)$ which exactly picks up the leading order in
$T$.  That rate function is the Legendre transform of the
log-generating function of the occupation and current statistics
which would give more direct access to the various cumulants, but
we will not need these here. Our ambition here is not so much on
the computational but rather on the conceptual level, to understand what
is the generic structure of $I(\mu,j)$ as well as its possible physical configuration.
In particular, we want  to stress
the similarities and the differences with Markov processes, and to
recognize the influence of modifying the waiting time
distribution.  As we will see explicitly in the examples of
section \ref{exa}, both the current and the occupation statistics
do pick up also higher moments of the
waiting time distribution, yielding markers for non-Markovian
behavior.

\subsection{Joint occupation-current statistics}\label{jointsec}
The fluctuation functionals appearing as rate functions such as
the $I(\mu,j)$ in the exponent of \eqref{dv}, are understood as
relative entropy densities, see \cite{var,DZ,el} for an
introduction to the systematic theory of large deviations.
The relative entropy is between a modified and the original
process where the modified process is chosen such as to make the
deviations typical. In our case, we deal with temporal processes
and the relative entropy density is like the rate of change of
dynamical entropies between the two processes. More specifically,
to compute $\bbP[\mu,j]:=\bbP[\mu_T \simeq \mu\,;\, j_T \simeq
j]$, we define a new semi-Markov process in the following way:
\begin{eqnarray*}
Q^*(x;\tau) &=& \frac{Q(x;\tau)\,e^{\Delta(x) \tau}}{Z(x)}\\
 p^*(x,y)&=&Z(x)\, p(x,y)\, e^{\frac{1}{2}F(x,y)}
\end{eqnarray*}
where $F(x,y)=-F(y,x)$, with  $\int_0^{\infty}Q^*(x;\tau)d\tau =1$
and $\sum_y p^*(x,y)=1$.  Comparing with \eqref{compare} we see
that $Q^*$ is of the form
\begin{eqnarray}\label{19}
 Q^*(x;\tau)p^*(x,y) &=& Q(x;\tau)p(x,y)e^{\frac{1}{2}F(x,y)+\Delta(x)\tau}\\
&=&
Q_0(x;\tau) p_0(x,y)\, e^{\frac{1}{2}(F(x,y)+G(x,y))+(\Delta^0_G(x)+\Delta(x))\tau}\nonumber
\end{eqnarray}
Hence, $\Delta^0_{F+G}(x) = \Delta^0_G(x)+\Delta(x)$. Most
important now, we require that $\mu$ and $j$ are stationary in
this new (modified) dynamics, i.e.,
\begin{equation}\label{c20}
 j(x,y) = \mu(x)\bar{\lambda}^*(x)p^*(x,y) - \mu(y)\bar{\lambda}^*(y)p^*(y,x)
\end{equation}
In terms of these new quantities the joint fluctuation functional
reads
\begin{equation}\label{imuj}
 I(\mu,j) = \sum_x \mu(x)\Delta(x) + \frac{1}{4}\sum_{x,y}j(x,y)F(x,y)
\end{equation}
Indeed, by using the explicit form~\eqref{ks} of the path-space measure we get
\begin{equation}
\bbP[\mu,j] = \int dP_{\mu}(\omega)\delta[\mu_T=\mu,j_T=j]  = \int
dP_{\mu}^{*}(\omega)\frac{dP_{\mu}}{dP^{*}_{\mu}}(\omega)\delta[\mu_T=\mu,j_T=j]
\end{equation}
where
\begin{eqnarray*}
 \log\frac{dP_{\mu}}{dP^{*}_{\mu}}(\omega) &\doteq&
 \sum_{i=1}^{n-1}\log
 \frac{Q(x_i;t_{i+1}-t_i)p(x_i,x_{i+1})}{Q^*(x_i;t_{i+1}-t_i)p^*(x_i,x_{i+1})}\\
&=& -\sum_{i=1}^{n-1}
\bigl[\frac{1}{2}F(x_i,x_{i+1})+\Delta(x_i) \bigr]\\
&=& -\frac{T}{4} \sum_{x,y} j_T(x,y)F(x,y)-T\sum_x\mu_T(x)\Delta(x)
\end{eqnarray*}
Note  that we have only written the extensive part in time,
because we are considering the large time limit anyway. Continuing
the computation, we now see that:
\begin{equation}
\bbP[\mu,j] =
 e^{-\frac{T}{4}\sum_{x,y}
 j(x,y)F(x,y)-T\sum_x\mu(x)\Delta(x)}\int dP_{\mu}^{*}(\omega)
 \delta[\mu_T=\mu,j_T=j]
\end{equation}
Finally,  in the large time limit we have that $\int
dP_{\mu}^{*,T}(\omega)\delta[\mu_T=\mu,j_T=j]\approx 1$, because
$\mu$ and $j$ are typical in the modified process. Therefore the
probability of the fluctuations has the asymptotic form:
\begin{equation}
 \bbP[\mu,j] \propto e^{-T I(\mu,j)}
\end{equation}
with $I(\mu,j)$ given in (\ref{imuj}).

\subsection{Occupation statistics}

A subquestion concerns the time-symmetric fluctuation sector; to
understand the statistics of the occupations {\it alone}.  That
means to look at \eqref{occ} and to write similarly to \eqref{dv},
\begin{equation}\label{dv2}
\bbP[p_T \simeq \mu] \propto e^{-T\,I(\mu)}
\end{equation}
To find the fluctuation functional $I(\mu)$ of only the occupation
statistics, one can perfectly repeat the argument for
occupation-current statistics above, but this time it suffices to
restrict oneself to the class of modified processes driven by gradient
forces, $F(x,y) = V(y) - V(x)$. Again, the point is that the potential
$V$ can be chosen such that it makes $\mu$ typical. The result for the
occupation fluctuation functional is
\begin{equation}
 I(\mu) = \sum_x \mu(x)\Delta(x)
\end{equation}
We see that  $\Delta$ appears as the quantity to average over with
$\mu$, for expressing the rate at which the system deviates from
the $\mu-$statistics. We can interpret $\Delta(x)$ as the excess
traffic of the modified process that makes $\mu,j$ typical, with
respect to the original process:
\begin{equation}
I(\mu) = \sum_x\mu(x)\Delta(x) = \dot{\caT}_{G+F}(\mu)-
\dot{\caT}_G(\mu)
\end{equation}
where the traffic functional is defined as
\begin{equation}\label{traf}
 \dot{\caT}_G(\mu) := \sum_x \mu(x)\Delta^0_G(x)
\end{equation}
to be compared with \eqref{fun}. Using (\ref{deltacond}), we can
deduce the response relation
\begin{equation}\label{cander}
 \frac{\partial\dot{\caT}_G(\mu)}{\partial G(x,y)} = -\frac{1}{2}\,j_{\mu,G}(x,y)
\end{equation}
with
\begin{equation}
 j_{\mu,G}(x,y) = \mu(x)\bar{\lambda}(x)p(x,y) - \mu(y)\bar{\lambda}(y)p(y,x)
\end{equation}
the expected transient current for a fluctuation $\mu$
in a dynamics as in \eqref{compare} determined by $G$. In this sense,
the traffic can be understood as a potential with respect to the currents;
 cf.\ similar remarks for the Markov processes, \cite{MN,mnw}. However, in
 contrast with the latter, the traffic~\eqref{traf} does not allow for a simple
 ``kinematic'' interpretation in terms of an expected number of jumps irrespectively
 of their direction.

\subsection{Fluctuation symmetry}
We now turn to the antisymmetric fluctuation sector, where we
recover the fluctuation theorem, cf.~\cite{ag,ell}. This is a
direct consequence of the local detailed balance we have proposed
in section \ref{sm}: Indeed, local detailed balance dictates that
(the extensive part of) the entropy production is
\begin{equation}
 S(\omega) \doteq \sum_{t\leq T}G(x_{t-},x_{t+}) = \frac{T}{2}\sum_{x,y}j_T(x,y)G(x,y)
\end{equation}
where $j_T$ is again the empirical current. As a consequence:
\begin{eqnarray*}
 \bbP[j_T \simeq j] &=& \int \id \bbP(\omega)\,\delta[j_T = j]\\
&=& \int \id \bbP(\theta\omega)\,e^{\frac{T}{2}\sum_{x,y}j_T(x,y)G(x,y)}\,\delta[j_T = j]\\
&=& e^{\frac{T}{2}\sum_{x,y}j(x,y)G(x,y)}\int \id
\bbP(\omega)\delta[j_T = -j]
\end{eqnarray*}
which means that
\begin{equation}\label{ft}
 \frac{ \bbP[j_T \simeq j]}{ \bbP[j_T \simeq -j]} \propto e^{\frac{T}{2}\sum_{x,y}j(x,y)G(x,y)}
\end{equation}
again asymptotically for $T \to +\infty$, i.e., up to temporal boundary terms.
Indeed and we already concluded in~\eqref{ldbref}, because of local detailed balance,
 $G(x,y)$ is the generalized thermodynamic force
for the transition $x \longrightarrow y$ in the entropy production.
Taking the logarithm and the limit $T\to +\infty$, formula \eqref{ft}
establishes a symmetry in the dynamical fluctuations of the entropy production, cf.~\cite{mnt}.

For convenience, we add in the next section some more details concerning the physical
interpretation of various players in the dynamical fluctuation theory, along with the
comparison to the Markov case.

\subsection{Comparison with Markov processes}

In the fluctuation functionals two new quantities appear:
$\Delta(x)$ and $F(x,y)$. The meaning of $F(x,y)$ can be made
clear through local detailed balance \eqref{ldb}. We already know
that for the original process, local detailed balance means that
\begin{equation}
 \frac{\bar{\lambda}(x)p(x,y)}{\bar{\lambda}(y)p(y,x)} = e^{\sigma(x,y)}
\end{equation}
where  $\sigma(x,y) = G(x,y) + u(x)-u(y)$ is the entropy flux
between system and reservoir per jump from $x$ to $y$, see
\eqref{ldbref}. For the modified process we find
 \begin{equation}
 \frac{\bar{\lambda}^*(x)p^*(x,y)}{{\lambda}^*(y)p^*(y,x)}
 = e^{\sigma(x,y)+\sigma_\text{ex}(x,y)}
\end{equation}
where  now $\sigma_\text{ex}(x,y)$ is the excess entropy flux of the
modified process with respect to the original process and is given
by
\begin{equation}\label{pote}
 \sigma_\text{ex}(x,y) =
 F(x,y)+\log\left(\frac{Z(x)\bar{\lambda}^*(x)}{\bar{\lambda}(x)}\right)-
 \log\left(\frac{Z(y)\bar{\lambda}^*(y)}{\bar{\lambda}(y)}\right)
\end{equation}
This means that, again up to some `potential difference,' (the last two terms in
\eqref{pote}) the term $F(x,y)$ is the extra force one adds to the
system to make $\mu$ and $j$ stationary. For a Markov process this
extra  `potential' becomes zero.

When the process is not Markov we
still have that
\begin{equation}
 \sum_{x,y}j(x,y)F(x,y) = \sum_{x,y}j(x,y)\sigma_{ex}(x,y)
\end{equation}
whenever the stationarity condition $\sum_y j(x,y) = 0$ is fulfilled.
In particular, this means that the potential terms do not make a time-extensive
contribution to the total entropy flux since along any trajectory, the empirical
 currents $j_T(x,y)$ always satisfy that stationarity condition up to corrections
 $O(1/T)$. In this sense, the second term in the joint fluctuation functional
$I(\mu,j)$ functional can really be called an excess entropy flux.\\

We have mentioned already above  how the quantity $\Delta(x)$
appears. In the case of a Markov process, $\Delta(x)$ is simply (minus) the
excess escape rate of the modified process with respect to the original process:
\begin{equation}\label{mc}
 \Delta(x) = \lambda(x)-\bar{\lambda^*}(x)
\end{equation}
Combining~\eqref{compare} with
\eqref{deltacond}, the
$\Delta^0_{G}(x)$ can be expanded around a Markov reference: e.g.,
by writing
\[
Q_0(x,y;t) = w(x,y)\,e^{-\lambda(x) t}\,\sum_{n=0}^{+\infty}
a^{(n)}(x)\,\frac{t^n}{n!}
\]
Then, the normalization \eqref{deltacond} leads to
\begin{eqnarray}\nonumber
\sum_y w(x,y)\,e^{\frac 1{2}G(x,y)} &=&
(\lambda(x)-\Delta^0_{G}(x))\,\Bigl(\sum_{n=0}^{+\infty}
\frac{a^{(n)}(x)}{(\lambda(x)-\Delta^0_{G}(x))^n}\Bigr)^{-1}\\
\sum_y w(x,y) &=& \lambda_x\,\Bigl(\sum_{n=0}^{+\infty}
\frac{a^{(n)}(x)}{\lambda_x^n}\Bigr)^{-1}\nonumber
\end{eqnarray}
Assuming that $a^{(0)}(x) = 1$ and that the other $a^{(n)}(x)$,
$n\geq 1$ are all small gives rise to an expansion of
$\Delta^0_{G}(x)$ and of $\Delta(x)$ around the Markov case~\eqref{mc}.

\section{Small fluctuations}\label{small}

In this section we
examine the regime of small (or Gaussian) fluctuations, to say more about
the traffic and the influence of the waiting time distributions on the
fluctuations. For this we make a quadratic approximation to the various
functionals but not (necessarily) around equilibrium.

\subsection{General}

We let the fluctuations in the empirical distribution and the empirical
current be parameterized with some $\epsilon$,
\begin{eqnarray*}
 \mu(x) &=& \rho(x)[1+\epsilon\mu_1(x)]\\
j(x,y) &=& j_{\rho}(x,y) + \epsilon j_1(x,y)
\end{eqnarray*}
where $\rho(x)$ is the stationary measure, and $j_{\rho}(x,y)$ is
the stationary current. Also $\Delta(x)$ and $F(x,y)$ are now of order
$\epsilon$ and to indicate that, we replace those with $\epsilon\Delta(x)$
and $\epsilon F(x,y)$. To first order in $\epsilon$, the conditions~\eqref{19}--\eqref{c20} become
\begin{align}\label{smallfluct}
 j_1(x,y) &=
 \frac{\rho(x)p(x,y)}{\left<\tau\right>_x}\Bigl[ \mu_1(x) +
 \Bigl(\frac{\left<\tau^2\right>_x-
 2\left<\tau\right>^2_{x}}{2\left<\tau\right>^2_x}\Bigr)
 \sum_zp(x,z)F(x,z)
 + \frac{1}{2}F(x,y) \Bigr] \nonumber
\\
 &\phantom{***}- `(x\longleftrightarrow y)'
\\ \label{smallfluct1}
 \Delta(x) &= -\frac{1}{2\left<\tau\right>_x}\sum_zp(x,z)F(x,z)
\end{align}
where
\begin{equation}
\left<\tau\right>_{x} = \frac{1}{\bar{\lambda}(x)}\,, \qquad
\left<\tau^2\right>_{x} = \int_0^{\infty}\id \tau\,
\tau^2\,Q(x;\tau)
\end{equation}
The fluctuation functional within the quadratic approximation is, cf.~\eqref{imuj},
\begin{equation}\label{I-quadratic}
 I(\mu,j) = \epsilon^2\sum_x\rho(x)\mu_1(x)\Delta(x)+\frac{\epsilon^2}{4}\sum_{x,y} j_1(x,y)F(x,y)
\end{equation}

We observe that for small fluctuations, only the first and second moments
of the waiting time distributions contribute. Furthermore, the second
term on the right hand side of~\eqref{smallfluct}) marks the
difference with the Markov case since for the exponentially distributed waiting times
$\left<\tau^2\right>_x = 2\left<\tau\right>^2_{x}$.
Due to the presence of this term beyond Markov, the functional
$I(\mu,j)$ no more splits into a sum over different transitions,
hence, it is responsible for the emergence of a new type of nonlocality
in the fluctuations, not present under the Markov condition.

\subsection{Close to equilibrium}

In the case of a (global) detailed balance dynamics, we can prove that
the occupation and current fluctuations become decoupled within the quadratic approximation.
Indeed, by using the detailed balance condition, the equation~\eqref{smallfluct}
can be explicitly solved for the extra forcing $F$:
\begin{equation}
\begin{split}
 F(x,y) &= \mu_1(y)-\mu_1(x) +  \Bigl(\frac{\left<\tau^2\right>_x
 -2\left<\tau\right>^2_{x}}{\left<\tau\right>_x}\Bigr)\Delta(x) - \Bigl(\frac{\left<\tau^2\right>_y
 -2\left<\tau\right>^2_{y}}{\left<\tau\right>_y}\Bigr)\Delta(y)
\\
 &\phantom{***}+ \frac{\left<\tau\right>_x}{\rho(x)p(x,y)}\,j_1(x,y)
\end{split}
\end{equation}
Substituted in the equation~\eqref{smallfluct1}, it yields
\begin{equation}
\Delta(x) =  -\frac{1}{2\left<\tau\right>_x}\sum_yp(x,y)
\Bigl[\mu_1(y)-\mu_1(x) +  \Bigl(\frac{\left<\tau^2\right>_x
-2\left<\tau\right>^2_{x}}{\left<\tau\right>_x}\Bigr)\Delta(x) - \Bigl(\frac{\left<\tau^2\right>_y
-2\left<\tau\right>^2_{y}}{\left<\tau\right>_y}\Bigr)\Delta(y)\Bigr]
\end{equation}
where we have used that $\sum_yj_1(x,y)=0$.
Clearly, that $\De$ only depends on the occupations $\mu_1$ and not on the currents.
That is why the occupation and the current fluctuations become statistically independent,
with the joint fluctuation functional being a sum of the occupation functional and
the current functional:
$I(\mu,j) = I(\mu) + I(j)$ where
\begin{align}\label{decoupled}
 I(\mu) &= \epsilon^2\sum_x\rho(x)\mu_1(x)\Delta(x)
\\ \intertext{and}
 I(j) &= \frac{\epsilon^2}{4}\sum_{x,y}
 \frac{\left<\tau\right>_x}{\rho(x)p(x,y)}\,j_1(x,y)^2\label{decoupled2}
\end{align}

Suppose now that the (global) detailed balance is slightly broken,
in the following sense: take $p(x,y)=p_0(x,y)+\epsilon p_1(x,y)$,
where $p_0(x,y)$ are detailed balanced transition probabilities.
As the functionals (\ref{decoupled})--(\ref{decoupled2}) are already of order $\epsilon^2$,
the small deviation from detailed balance will not contribute. We have
therefore still uncorrelated statistics for the occupations and currents
in the close-to-equilibrium regime, and the marginal fluctuation
functionals~\eqref{decoupled}--(\ref{decoupled2}) remain unchanged.

Remark that whereas the current statistics has exactly the same form as
 for a Markov process with escape rates $\lambda(x) = 1 / \left<\tau\right>_x$,
 we observe a difference in the occupation statistics. As the difference between
semi-Markov and Markov processes lies in the time-symmetric part of the path-space
 probabilities, it should come as no surprise that the occupation statistics are
 more sensitive to details of the waiting time distribution than are the current statistics.

\section{Examples}\label{exa}

First we give an example of semi-Markov process obtained from a Markov model via coarse-graining.
Its generalized version is then used to illustrate our dynamical fluctuation theory.

\subsection{Semi-Markov from Markov}
\label{ex1}

\begin{figure}[ht]
\center{\includegraphics[scale=0.2]{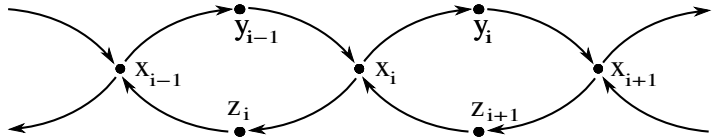}}
\caption{Markov random walk with `hidden' states.}
\label{picture1}
\end{figure}

Consider a Markov random walk as in~Fig.~\ref{picture1}, with three kinds of states $x_i, y_i$, and
$z_i$ for $i=1,\ldots, N$ on a ring ($N+1 \equiv 1$).

As the arrows in Fig.~\ref{picture1} suggest, the only transitions
allowed are $x_i\to y_i$, $x_i\to z_{i}$, $y_i\to x_{i+1}$, and
$z_i\to x_{i-1}$, whereas all the others are are forbidden. This is
a model of a one-dimensional random walk with `hidden' states: the particle at state $x_i$ can 
go `to the right' (from $x_i$ to $x_{i+1}$)
 through the `hidden' state $y_i$ only, and `to the left' through $z_i$ only.
  More specifically, we fix $\lambda_x, \lambda_y, \la_z > 0$ and we set the
Markov transition rates to
\begin{eqnarray*}
W(x_i\to y_i) &=& p \lambda_x\\
W(x_i\to z_{i}) &=& q \lambda_x\\
W(y_i\to x_{i+1}) &=& \lambda_y\\
W(z_i\to x_{i-1}) &=& \lambda_z
\end{eqnarray*}
for some $p+q=1$. The rates do not depend on the position $i$ on the ring.

\begin{figure}[ht]
\center{\includegraphics[scale=0.2]{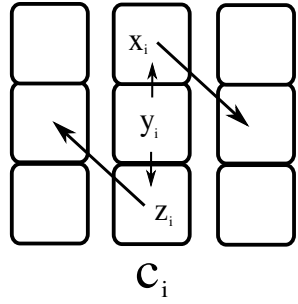}}
\caption{Coarse-graining of the states.}
\label{picture2}
\end{figure}

Since every pair of `hidden' states $y_i$ and $z_i$ have the unique precursor
$x_i$, we can follow a simple coarse-graining procedure:
for each $i$ to take together all three states $x_i$, $y_i$, and $z_i$,
see Fig.~\ref{picture2}. By construction, the coarse-grained random walk
on the new `block' states, denoted by $c_i$, is a semi-Markov process.

We can calculate the probability that the
walker occupies $c_i$ for a time $\tau > t$ before jumping to
$c_{i+1}$, and its time derivative is the density
$Q(c_i,c_{i+1};t)$. Exploiting that the only possibility of going from
$c_i$ to $c_{i+1}$ (or $c_{i-1}$) is via the `hidden' state $y_i$ (or $z_i$),
the waiting time distributions~\eqref{defs} read
\begin{eqnarray*}
Q(c_i,c_{i+1};t) &=& \int_0^{t} \id\tau\, p\,\lambda_x
e^{-\lambda_x\tau}\cdot\lambda_y\,e^{-\lambda_y(t-\tau)}\\
&=& p\, \lambda_x\, \lambda_y\frac{e^{-\lambda_y
t} - e^{-\lambda_x t}}{\lambda_x-\lambda_y}\\
Q(c_i,c_{i-1};t) &=& q\,\lambda_x\,\lambda_z \frac{e^{-\lambda_z
t} - e^{-\lambda_x t}}{\lambda_x-\lambda_z}
\end{eqnarray*}
It is clear that $Q(c_i,c_{i+1};t)$ and $Q(c_i,c_{i-1};t)$
determine the dynamics of the new stochastic process which is
semi-Markov: the updates are decided by the immediate history but
the waiting time distribution is not exponential and it can depend
on the specific transition. Observe also that
\begin{eqnarray*}
\int_0^{+\infty} \id t\,Q(c_i,c_{i+1};t) = p\,,\qquad
\int_0^{+\infty} \id t\,Q(c_i,c_{i-1};t) = q
\end{eqnarray*}
which indicates a driving whenever $p\neq q$.  In all events the
stationary distribution is uniform over the ring.\\

Let us investigate two different limits of this example. First let
$\lambda_x\to +\infty$. In this limit we get
\begin{eqnarray}\label{1ex}
Q(c_i,c_{i+1};t) = p\lambda_y\,e^{-\lambda_y
t}\,,\qquad
Q(c_i,c_{i-1};t) = q\lambda_z\,e^{-\lambda_z t}
\end{eqnarray}
Unless $\la_y = \la_z$, the process is not Markov.
In general it is semi-Markov with direction dependent waiting time
distribution.

Secondly we consider the limits
$\lambda_y\to\lambda_x$ and $\lambda_z\to\lambda_x$ together. The resulting
transition densities are
\begin{eqnarray}\label{2ex}
Q(c_i,c_{i+1};t) = p\, \lambda_x^2\; t\,e^{-\lambda_x
t}\,,\qquad
Q(c_i,c_{i-1};t) = q\,\lambda_x^2 \;t\,e^{-\lambda_x t}
\end{eqnarray}
Here the time dependence of the two clocks for going to the left
or to the right are the same. In particular, it does not matter
whether the walker first picks a direction or just picks the first
clock that rings. Still it is not a Markov process, because the
waiting time distributions of the clocks are not exponential. This
process is semi-Markov with waiting time-direction
independence.

\subsection{CTRW on the ring}

The semi-Markov model obtained in the previous section is an example of
continuous time random walk (CTRW). In this section we give some explicit
solutions to equations that appear in dynamical fluctuation theory, for a
CTRW on the ring.

We continue
with states that represent the sites on a ring of length $N$, with
translation invariance as in the above explicit example. Let the transition densities be
\[
 Q(x,\tau) = Q(\tau), \ \ \ \ \ \ p(x,x+1)= p,\ \ \ \ \ \
  p(x+1,x) = q
\]
We restrict to the small fluctuations as in Section~\ref{small}.
Because the current fluctuation $j=j_{\rho}+\epsilon j_1$ has to satisfy
$\sum_y j(x,y)=0$, we see that $j_1(x,x+1)+j_1(x,x-1) = 0$, and
therefore
$j_1(x,x+1) = j_1$ is a constant on the ring.

The equations (\ref{smallfluct}) become
\begin{eqnarray}\label{smallfluctring}
j_1 &=& \frac{1}{N\left<\tau\right>}
\left[ p\mu_1(x)-q\mu_1(x+1) -AF(x+1,x+2) +BF(x,x+1)-AF(x-1,x)\right]\nonumber\\
\Delta(x) &=& \frac{1}{2\left<\tau\right>}(qF(x-1,x)-pF(x,x+1))
\end{eqnarray}
where the constants $A$ and $B$ are
\begin{eqnarray*}
A &=& p\,q \Bigl(
\frac{\left<\tau^2\right>}{2\left<\tau\right>^2} - 1 \Bigr)\\
B &=& \frac{(p^2+q^2)\left<\tau^2\right>}{2\left<\tau\right>^2}-\frac{(p-q)^2}{2}
\end{eqnarray*}
The fluctuation functional reads
\begin{equation}\label{ff}
 I(\mu,j) = \frac{\epsilon^2}{N}\sum_x\mu_1(x)\Delta(x) + \frac{\epsilon^2j_1}{2}\sum_xF(x,x+1)
\end{equation}
The second term on the right-hand side is easily  computed by summing~\eqref{smallfluctring} over all $x$.
Using that $\sum_x\mu_1(x) = 0$ we get
\[
 \sum_xF(x,x+1) = \frac{N^2\left<\tau\right>j_1}{B-2A}
\]
Defining now $f(x,x+1)$ by $F(x,x+1) = f(x,x+1)+
\frac{N\left<\tau\right>j_1}{B-2A}$, we see that
$\sum_xf(x,x+1)=0$, hence, $f$ is of the gradient form
$f(x,x+1) = V(x+1) - V(x)$. Substituting in~\eqref{smallfluctring}, we get
\begin{eqnarray}\label{smallimu}
0 &=& \frac{1}{N\left<\tau\right>} \left[ p\mu_1(x)-q\mu_1(x+1) -
Af(x+1,x+2) +Bf(x,x+1)-Af(x-1,x)\right]\nonumber\\
\Delta(x) &=&
\frac{1}{2\left<\tau\right>}(qf(x-1,x)-pf(x,x+1)) +
(q-p)\frac{N^2\left<\tau\right>j_1}{B-2A} \nonumber\\
&=:& \Delta'(x) + (q-p)\frac{N^2\left<\tau\right>j_1}{B-2A}
\end{eqnarray}
Finally, \eqref{ff} takes the form
\begin{eqnarray}
 I(\mu,j) &=& I(\mu) + I(j) \nonumber\\
&=&\frac{\epsilon^2}{N}\sum_x\mu_1(x)\Delta'(x)+
\frac{\epsilon^2N^2\left<\tau\right>j_1^2}{2(B-2A)}
\end{eqnarray}
So there is a decoupling of current and occupation statistics.
Whereas, in general, this occurs only for small fluctuations and close to
equilibrium, here it is apparently valid arbitrarily far from equilibrium.
The fundamental reason lies in the translation-invariance property of the dynamics.\\

Let us further examine the statistics of current fluctuations,
exploiting that the functional $I(j)$ is explicit.
Using that in the Markov case $2(B-2A)=1$, we can write in general
$I(j) = C^{-1}I_M(j)$, with $I_M(j)$ the fluctuation functional for a
Markov process with the same average waiting time
$\left<\tau\right>$, and with $C$ given by
\begin{equation}
 C  = 1+
 (2p-1)^2\left(\frac{\mbox{Var}(\tau)}{\left<\tau\right>^2}-1\right)
\end{equation}
where $\text{Var}(\tau) =
\left<\tau^2\right>-\left<\tau\right>^2$ is the variance of the waiting
time distribution. Thus $C$ is a correction
factor with respect to the Markov case; one checks that $C \geq 0$.
We also see that the bigger the variance of the waiting
times, the smaller $C^{-1}$, and therefore the $I(j)$ will become
flatter. Remark that in detailed balance (i.e.\ for $p=\frac{1}{2}$),
 we have that $C=1$. So in this case one has the same fluctuation
functional as in the Markov case, as it should be according to~\eqref{decoupled}.
 Furthermore, there is a fundamental difference between the cases
$\text{Var}(\tau) > \left<\tau\right>^2$ and $\text{Var}(\tau) < \left<\tau\right>^2$
 (i.e., whether the variance of the waiting times is bigger or smaller than
in the Markov case). When the variance is smaller than in the Markov case,
$C$ as a function of $p$ tends to get bigger when $p$ gets closer to $0$ or $1$.
 In the other case, $C$ becomes smaller when $p$ is closer to $0$ or $1$.\\

Finally we can compute the occupation statistics $I(\mu)$.
Formul{\ae} considerably complicate however when $N$ is large and
that is why we choose to be explicit only for $N=3$ (a ring with three sites).
 Similarly as for $I(j)$ we can write
$I(\mu) = C'^{-1}I_M(\mu)$, with
\begin{eqnarray*}
 I_M(\mu) &=& \frac{\epsilon^2}{3\left<\tau\right>}\sum_x(p\mu_1(x)-q\mu_1(x-1))^2\\
&=& \frac{2\epsilon^2}{3\left<\tau\right>}(1+p^2-p)[\mu_1(1)^2+\mu_1(2)^2-\mu_1(1)\mu_1(2)]
\end{eqnarray*}
using that $\mu_1(1)+\mu_1(2)+\mu_1(3)=0$. The $C'$ is given by
\begin{equation}
 C' = 1+(p^2-p+1)\Bigl(\frac{\mbox{Var}(\tau)}{\left<\tau\right>^2}-1\Bigr)
\end{equation}
Also here, the bigger the variance of the waiting times, the flatter the
fluctuation functional becomes. However, in the detailed balance case the
functional does not reduce to the Markov-form. This is indeed what we concluded
in the discussion after (\ref{decoupled}). There is another difference with
the current fluctuations: as a function of $p$, the fluctuation functional
gets flatter when $p$ gets closer to $0$ or $1$, independent of the sign of
 $\mbox{Var}(\tau) / \left<\tau\right>^2-1$.

\subsection{Generating function approach to current statistics}

Finally we consider (arbitrary) current fluctuations on the
ring and compare two possible approaches. As explained in the present paper,
 one way of computing this is via contraction of the
joint fluctuation functional:
\begin{equation}
 I(j) = \inf_{\mu}I(\mu,j)
\end{equation}
As is however often the case, explicit computations proceed more
easily via the generating function
\begin{equation}
 G(v) = \left<e^{Tvj_T}\right>
\end{equation}
Defining $g(v) = \lim_{T\to\infty}\frac{1}{T}\log G(v)$, one can
prove that $g(v)$ exists and is the Legendre transform of the
fluctuation functional $I(j)$:
\begin{equation}
 g(v) = \sup_j\{vj-I(j)\}
\end{equation}
and vice versa.
It generates the current cumulants, see e.g.~\cite{bo,der,ba} for
applications to nonequilibrium interacting particle systems. By
using the Laplace transform as in \eqref{laplace}, with
$\tilde{Q}(s)$ the Laplace transform of $Q(\tau)$, we solve the
equation
\begin{equation}\label{geneq}
 \tilde{Q}(s^*) = \frac{1}{pe^{v}+qe^{-v}}
\end{equation}
for $s^*$; then $g(v) = s^*$.

As an example, consider the waiting time distribution
\begin{equation}\label{ga}
 Q(\tau) = \frac{1}{\Gamma(a)}\tau^{a-1}\lambda^ae^{-\lambda \tau}
\end{equation}
for $\lambda >0, a\geq 1$. (Note that $a=1$ represents the Markov
case.) For this distribution,
\[
 m = \left<\tau\right> = \frac{a}{\lambda}\ \ \ \ \ \ \sigma^2 = \mbox{Var}(\tau)
  = \frac{a}{\lambda^2}
\]
and (\ref{geneq}) becomes
\begin{equation}
 \tilde{Q}(s^*) = \left(\frac{\lambda}{\lambda+s^*}\right)^a = (pe^{v}+qe^{-v})^{-1}
\end{equation}
or,
\begin{equation}
 g(v) = \lambda(pe^{v}+qe^{-v})^{\frac{1}{a}} - \lambda
\end{equation}
By taking derivatives at $v=0$ we obtain the next explicit expressions for the current moments:
\begin{eqnarray*}
 j_{\rho} &=& \frac{p-q}{m}\\
 \frac{1}{T} \left<(j_T-j_{\rho})^2\right> &=& \frac{1}{m}(\frac{\sigma^2}{m^2}-1)(p-q)^2 +\frac{1}{m}\\
 \frac{1}{T} \left<(j_T-j_{\rho})^3\right> &=&
 \frac{1}{m}(\frac{\sigma^2}{m^2}-1)(\frac{\sigma^2}{m^2}-2)(p-q)^3
+ \frac{1}{m}(\frac{3\sigma^2}{m^2}-2)(p-q)
\end{eqnarray*}
It is interesting to see that the variance of
the currents picks up a term depending on the driving $p-q$ and that this
term is only non-zero if the process is non-Markov. This is indeed what we
expect from the discussions about the small fluctuations regime, where we
found that in, or close to, the detailed balance case the current fluctuations
are quite insensitive to changes in the variance of the waiting times.

\begin{acknowledgments}
K.~N.\ acknowledges the support from the project
AV0Z10100520 in the Academy of Sciences of the Czech Republic
and from the Grant Agency of the Czech Republic (Grant no.~202/07/J051). C.M.
benefits from the Belgian Interuniversity Attraction Poles Programme P6/02. B.~W.\ is an aspirant of FWO Flanders.
\end{acknowledgments}

\appendix

\section{More background on semi-Markov processes}\label{apA}

We give some more details on the structure of semi-Markov
processes. See \cite{mon,qw} for excellent introductions.

\subsection{Embedded Markov chain}
To have a good intuition about a semi-Markov process it is worth
seeing it as a discrete time Markov chain to which specific
waiting times are added. In particular, the discrete sequence of
states in the semi-Markov process is drawn from the Markov chain
with transition probabilities $p(x,y)$: if at time $\tau_0$ the
state $x_0$ was created, then we choose the next state $x_1$ with
probability $p(x_0,x_1)$. Secondly, we have to decide how long has
been the waiting between the creation of $x_0$ and the creation of
$x_1$. For that we need the waiting time distribution: let its distribution
(i.e. the integrated probability density)
be $1-\frac{\Lambda(x,y;t)}{p(x,y)}$, so that the (total)
distribution function for the random waiting time $\tau_1-\tau_0$ between
the creation of $x_0$ and the creation of $x_1$,
is $p(x_0,x_1)-\Lambda(x_0,x_1;t)$.

A semi-Markov process is thus constructed most elegantly from a
Markov renewal process $(x_n,\tau_n)_{n\geq 0}$ for $x_n\in
\Omega$ and $\tau_0\leq \tau_1\leq\ldots\leq \tau_n\leq\ldots$
denoting jump times, with transition probabilities
\[
\bbP[x_{n+1}=y,\tau_{n+1} -\tau_n\leq t \rel x_n,\tau_n,
x_{n-1},\tau_{n-1},\ldots,x_0,\tau_0] = p(x_n,y)-\Lambda(x_n,y;t),
\quad n, t\geq 0
 \]
 satisfying all of \eqref{defs} with $Q(x,y;t) = - \frac{\id}{\id t}\Lambda(x,y;t)$.\\
The process $(x_n)$ with transition probabilities $p(x,y)$ is
called the embedded Markov chain. For the randomness in time, we
define the jump counting process $\nu(t), t\geq0$ of the total
dynamical activity up to time $t$,
\[
\nu(t) = \sup\{n\geq 0: \tau_n\leq t\}
\]
which we assume is finite with probability one. The semi-Markov
process corresponding to that renewal process is then defined by
\[
x(t) = x_{\nu(t)},\quad t\geq 0
\]
where the state at time $t$ is just equal to $x_n$ of the embedded
Markov chain, if $\nu(t)=n$. For a Markov process $\nu(t)$ is just
a Poisson process.

\subsection{Master equation}
An intuitive way of deriving a generalization of the master
equation for
 semi-Markov processes is by considering the corresponding Markov
 process $(x_t,\tau_t)_{0\leq t\leq T}$, where $x_t$ is the
 configuration of the semi-Markov process at time $t$ and $\tau_t$ is the time
 that the system has been in this configuration since its last jump.\\

With this in mind we can write down the transition probabilities
for  the process $(x_t,\tau_t)$,
\begin{equation}
 P(x,\tau;y,\tau') = \mbox{Prob}(x_{t+\id t}=y,\tau_{t+\id t}=\tau'|x_{t}=x,\tau_{t}=\tau)
\end{equation}
We then have that with $\lambda(x;\tau)=Q(x;\tau)/\Lambda(x;\tau),
\Lambda(x;\tau) = \sum_y\Lambda(x,y;\tau), \lambda(x,y;\tau) =
Q(x,y;\tau)/\Lambda(x,y;\tau)$,
\begin{itemize}
 \item[i)]
 $P(x,\tau;x,\tau+\id t) = 1-\lambda(x,\tau)\id t + o(\id t)$;
 \item[ii)]
 $P(x,\tau;y,0) = \lambda(x,y;\tau)\id t + o(\id t)$.
\end{itemize}
and other transition probabilities are of order $o(\id t)$.

Let us now look at the evolution  of probability densities
$\mu_t(x,\tau$) in this dynamics. It is easily seen that for
$\tau\neq 0$:
\begin{equation}\label{discrete}
 \mu_{t+\id t}(x,\tau) = \mu_t(x,\tau-\id t)[1-\id t\,\lambda(x;\tau-\id t)]
\end{equation}
from which it follows that
\begin{equation}\label{mastertau}
 \frac{\partial \mu_t(x,\tau)}{\partial t} =
 -\frac{\partial \mu_t(x,\tau)}{\partial \tau} - \mu_t(x,\tau)\lambda(x;\tau)
\end{equation}
and for $\tau=0$,
\begin{equation}\label{tauzero}
 \mu_{t}(x,0) = \int_{0}^{\infty}\id\tau\sum_y\mu_t(y,\tau)\lambda(y,x;\tau)
\end{equation}
Because we are mainly interested in the process $x_t$ (which is no
longer Markov), we integrate over the waiting times $\tau$:
$\mu_t(x) = \int_{0}^{\infty}\id\tau\mu_{t}(x,\tau)$. Doing that
for \eqref{mastertau} we get (assuming that $\mu_t(x,\infty)=0$):
\begin{equation}
 \frac{\partial \mu_t(x)}{\partial t} =
 \mu_t(x,0) - \int_0^{\infty}\mu_t(x,\tau)\lambda(x;\tau)\,\id
 \tau
\end{equation}
Using (\ref{tauzero}) we thus arrive at a generalized Master
equation:
\begin{equation}\label{genm}
 \frac{d\mu_t(x)}{dt} = \int_0^{\infty}d\tau\sum_y
 \left[ \mu_t(y,\tau)\lambda(y,x;\tau)-
 \mu_t(x,\tau)\lambda(x,y;\tau)\right] = -\sum_yj_t(x,y)
\end{equation}
which also defines the currents $j_\mu(x,y)$.

There is obviously a downside to this equation, and that is that
we need knowledge of $\mu_t(x,\tau)$ to compute the time
derivative of $\mu_t(x)$. There is a way to circumvent this
problem by using the  Laplace transform.\\

Taking the Laplace transform of a function $f$ as
\begin{equation}\label{laplace}
 \tilde{f}(s) = \int_0^{\infty}f(t)e^{-st}\,\id t
\end{equation}
and putting $\mu_0(x,\tau) =
\mu_0(x)\bar{\lambda}(x)\Lambda(x,\tau)$ it can be proven that
\[
 \int_0^{\infty}\id \tau\mu_T(x,\tau)\lambda(x,y;\tau) =
 \int_0^T \mu_t(x)\,\varphi(x,y;T-t)\id t + \mu_0(x)\left(\bar{\lambda}(x)p(x,y)-
 \int_0^T \id t\varphi(x,y;t)\right)
\]
with $\varphi$ having Laplace transform
\[
 \tilde{\varphi}(x,y;s) = \frac{\tilde{Q}(x,y;s)}{\tilde{\Lambda}(x;s)}
\]
We substitute that in the master equation:
\begin{eqnarray}
 \frac{d\mu_T(x)}{dT}&=&\sum_y\int_0^T \id t\left\{ \left[\mu_t(y)-
 \mu_0(y)\right]\varphi(y,x;T-t)
 - \left[\mu_t(x)-\mu_0(x)\right]\varphi(x,y;T-t) \right\} \nonumber\\
 &&+ \sum_y\left[ \mu_0(y)\bar{\lambda}(y)p(y,x) - \mu_0(x)\bar{\lambda}(x)p(x,y) \label{malap}\right]
\end{eqnarray}
The function $\varphi$ represents the memory of the semi-Markov
process. The faster $\varphi(x,y;t)$ decays to zero, the less
memory we have. For example, for a Markov process with escape
rates $\lambda(x)$ we find that $\varphi(x,y;t) =
\lambda(x)p(x,y)\,\delta(t)$. Indeed a Markov process has no
memory (and the master equation for Markov processes arises as a
special case of this generalized equation).

\subsection{Stationarity}\label{sta}

Stationarity  of the semi-Markov process means that path-space
averages of time-independent observables are still time
independent. But as above we can use the corresponding Markov
process $(x_t,\tau_t)_{0\leq t\leq T}$. Stationarity of the
semi-Markov process is ensured by demanding that
$\mu_t(x,\tau):=\rho(x,\tau)$ is stationary. Solving then
(\ref{mastertau}) with the LHS zero, and using $\rho(x) =
\int_0^{\infty}d\tau\rho(x,\tau)$, we get $\rho(x,\tau) =
\rho(x)\bar{\lambda}(x)\Lambda(x;\tau)$. This means that if we
know that the system is in a configuration $x$, then the
probability that it has been there already for a time
$\tau$ is equal to $\bar{\lambda}(x)\Lambda(x;\tau)$.

The stationary measure of our semi-Markov process is $\rho(x)
\propto \frac{\pi(x)}{\bar{\lambda}(x)}$ where $\pi(x)$ is the
stationary measure of the embedded Markov chain.  We write
\[ \rho(x) =
\frac{1}{\xi}\frac{\pi(x)}{\bar{\lambda}(x)}\] for some
normalization $\xi$, the overall average waiting time. In this
notation, we see that the stationary currents are
\begin{equation}
 j_{\rho}(x,y)=\frac{\pi(x)p(x,y) - \pi(y)p(y,x)}{\xi}
\end{equation}
and that they are zero iff the embedded Markov chain is detailed
balance. That however is only equivalent with time-reversal
invariance if the semi-Markov process is time-direction
independent, see e.g. \cite{cha,qw,wq}.



\begin{thebibliography}{10}
\bibitem{remark1}
There are always infinitely many reference equilibrium processes
such that this relation is true. Its satisfaction makes the coming
large deviation argument particularly convenient since the stochastic
modification needed to make a fluctuation typical will appear to have
an identical form, see Section \ref{jointsec}.
\end{thebibliography}
\end{document}